\documentstyle[aps]{revtex}
\begin{document}
\author{A.V.Dooglav$^{1,2}$, A.V.Egorov$^1$, I.R.Mukhamedshin$^1$, A.V.Savinkov$^1$,
H.Alloul$^2$, J.Bobroff$^2$, W.A.MacFarlane$^2$, P.Mendels$^2$, G. Collin$^3$%
, N. Blanchard$^3$, P.G.Picard$^3$}
\address{$^1${\it Magnetic Resonance Laboratory, Kazan State University, 420008}\\
Kazan, Russia \\
$^2${\it Laboratoire de Physique des Solides, Universite de Paris-Sud,} {\it %
91405 Orsay, France}\\
$^3${\it Laboratoire de Leon Brillouin, CE Saclay, CEA-CNRS, 91191}\\
Gif-sur-Yvette, France}
\title{Antiferromagnetism in hydrated 123 compounds}
\date{The Date }
\maketitle

\begin{abstract}
Copper nuclear quadrupole resonance (NQR) and zero field nuclear magnetic
resonance (ZFNMR) studies of YBa$_2$Cu$_3$O$_{6.5}$ show that a magnetic
phase appears in underdoped 123 superconductors treated in ambient wet air.
The studies give convincing evidence that the ``empty'' CuO chains play the
role of easy water insertion channels. The reaction occurs first in ordered
regions of the crystallites. The final product of the reaction is a
non-superconducting antiferromagnetic compound characterized by at least two
types of magnetically ordered copper ions with ZFNMR spectra respectively in
the frequency ranges of 46-96 and 96-135 MHz respectively. Even for powder
samples fixed in an epoxy resin, the reaction is found to occur partly after
a few years.\\
\end{abstract}


The idea of co-existence of superconductivity and local antiferromagnetism
in high-temperature superconductors (HTSC) is appearing more and more
frequently in theoretical and experimental papers. Recently we found \cite
{ref1} that in aged samples of YBa$_2$Cu$_3$O$_{6+x}$ with $x<0.8$ packed in
paraffin and kept at room temperature for about 6 years the superconducting
volume fraction had essentially decreased. In the same samples, we also
found, along with the typical copper NQR spectrum of the superconducting
1-2-3 system, copper ZFNMR spectra in the frequency range 46-135 MHz,
indicating the presence of a magnetically ordered phase which remains
observable up to 200 K.

The initial goal of this study was to find a way of artificially aging the
compound to produce the magnetic phase. At the time of our previous report 
\cite{ref1}, we considered the presence of the magnetic phase as a result of
stripe pinning due to the Ortho-III phase, which was well-ordered during the
very long-term room temperature annealing, so we began the artificial aging
experiments with attempts to accelerate the oxygen ordering process. It is
known \cite{ref2} that the temperature boundary of the Ortho-III phase is
about 75$^o$C, so the first experiments were devoted to annealing at 55-60$%
^o $C for a few weeks, but this failed to produce the magnetic phase. Thus,
further attempts were undertaken at somewhat higher temperatures. Finally we
managed to get a small amount of magnetic phase exhibiting the
characteristic ZFNMR spectra \cite{ref1}. However, the mass of the samples
during the ambient air annealing had increased, and controlled annealing of
dry powder sealed in an ampoule showed neither the mass changes nor the
appearance of the magnetic phase.

The most natural explanation of the mass increase is the occurrence of a
chemical reaction of YBCO with components of the air, in particular, with
water vapour, since the significant reactivity of 123 with water is well
established. Most of the early papers on the reaction of 123 with water
reported reaction products such as Y$_2$BaCuO$_5$ (``green phase''), CuO,
BaCO$_3$ and some others. Some of these compounds are antiferromagnets
having either low N\'eel temperature (15-30 K) or ZFNMR spectra very
different from the one we observed in Ref.\cite{ref1}. More recent studies
(see Ref.\cite{ref3} and references therein) using high-resolution electron
microscopy and X-Rays have shown that the 123 superconductors react with
water vapour via a topotactic mechanism. The final product of this reaction,
occuring in a bulk material at 75-250$^o$C, is a so-called pseudo-248 phase H%
$_{2z}$YBa$_2$Cu$_3$O$_{6+x+z}$ (isostructural with the familiar 248
structure but with 50\% of the Cu(1) sites vacant and referred to as phase $B
$ in Ref.\cite{ref3}).

To study the changes occuring in 123 HTSC under reaction with water vapour
by NQR and NMR a number of samples with different water content were
prepared. As starting material, we used samples of YBa$_2$Cu$_3$O$_{6.5}$ (T$%
_c$=56 K) in the form of free powder with a particle size of about 30 $\mu $%
m, freshly synthesized using conventional solid state reaction of powdered Y$%
_2$O$_3$, BaO and CuO at about 940$^o$C interrupted periodically for
grinding. The free powder annealing was done in air at 150$^o$C. A vapour
pressure (about 35 mbar at 150$^o$C) was established in a closed furnace by
evapouration. The samples were weighed periodically to measure the water
uptake. After annealing all samples were packed in Stycast 1266A epoxy. With
this procedure, we obtained a number of samples of YBa$_2$Cu$_3$O$_{6.5}$(H$%
_2$O)$_z$ with water concentrations z=0; 0.14; 0.24; 0.55 and 1.2, the last
value being the maximum water uptake, beyond which further annealing did not
yield a mass increase.

Home-built pulsed NMR spectrometers were used to measure NQR and ZFNMR
spectra at 4.2 K. The copper NQR spectra of the free-powder annealed samples
are shown in Fig.1. Since the nuclear spin-lattice relaxation rate of the
``chain'' copper is significantly slower than that of the ``plane'' copper 
\cite{ref1}, it is possiple to separate their spectra (Fig.1b,c). The large
difference in their transverse relaxation rates can also be used for
separation. The assignment of different NQR lines to the nuclei of Cu(1)
residing in different oxygen coordinations is done in Ref. \cite{ref5}. The
fast relaxing part of the copper NQR spectrum (Fig.1c) is usually assigned
to the plane copper nuclei. It consists of two groups of lines at
frequencies 30.8 and 28 MHz for the isotope $^{63}$Cu(2). To simplify the
discussion, hereafter we will discuss only the frequencies and intensities
of the NQR lines for this isotope; naturally, the corresponding $^{65}$Cu
lines are also observed.

One can see in Fig.1 that the intensity of the copper NQR spectrum decreases
with increasing water uptake. It is remarkable that at the initial stage of
the reaction (at a small water uptake) the narrow $^{63}$Cu NQR lines at
31.5 and 31 MHz, i.e. the lines of $^{63}$Cu$^{+}$ nuclei belonging to the
``empty'' ...-Cu-Cu-... chains located between two ``full'' ...-Cu-O-Cu-...
chains (31.5 MHz, Ortho-II phase) and between ``full'' and ``empty'' chains 
\cite{ref5}, begin to disappear first. At the same time the broad
fast-relaxing line at 28 MHz (the plane $^{63}$Cu$^{2+}$(2)) also dereases.
At water uptake of approximately 0.5 molecules of H$_2$O per YBa$_2$Cu$_3$O$%
_{6.5}$, the narrow lines of Cu(1) and the Cu(2) line at 28 MHz are
practically absent in the NQR spectrum, and the intensity of the NQR line of
Cu(2) at 30.8 MHz has decreased by a factor of 2 compared to the unreacted
compound. The intensity of the NQR line belonging to the triply-coordinated
copper nuclei (located at the ends of the chains, 24 MHz) doesn't change
significantly, though a small increase of its intensity at the first stages
of water uptake was noted. At maximum water uptake (1.2 molecules) the
copper NQR spectrum typical for YBa$_2$Cu$_3$O$_{6.5}$ has disappeared,
giving way to a broad copper NQR line at 18.4 MHz.

The copper ZFNMR spectra of free-powder annealed samples are shown in Fig.2.
At small water uptake ($z<0.5$) the spectrum, as in Ref.\cite{ref1},
consists of two groups of well-resolved lines. The low-frequency group
(46-96 MHz) resembles the Cu(2) ZFNMR spectrum of antiferromagnetic PrBa$_2$%
Cu$_3$O$_7$\cite{ref6} and corresponds to a hyperfine magnetic field of $%
\sim $64 kOe at the copper nucleus and quadrupolar frequency of $\nu
_Q\approx 30$ MHz. A field of about 103 kOe and $\nu _Q\approx 16$ MHz
produces a high-frequency group (96-135MHz), this part of the spectrum is
practically identical to that of the Nd$_2$CuO$_4$ antiferromagnet \cite
{ref7}. The relative intensity of the two groups corrected for the
difference in transverse relaxation rate and the (square) frequency
dependence is estimated to be $I_{LFG}/I_{HFG}=1\pm 0.5$.

The intensities of both groups of the spectrum scale with water uptake for $%
z<0.5$ (Fig.2b), while for $z>0.5$ a new component appears and grows to
maximal intensity in the sample with maximum water uptake. Although the
separate lines of this component are not well-resolved, it can be assigned
to copper nuclei located in a hyperfine field of approximately 80 kOe, just
like in YBa$_2$Cu$_3$O$_6$ compound.

Changes in the superconducting volume fraction were determined by measuring $%
ac$ diamagnetic susceptibility ($H_1\approx 1$ Oe, $f=1$ kHz). The
superconducting volume fraction decreases with water uptake (Fig.3) while T$%
_c$ remains practically unchanged. The sample with maximum water uptake is
completely non-superconducting.

So, in accordance with \cite{ref3}, we conclude that insertion of water into
a 123 compound proceeds in two stages. The first stage ($z<0.5$) is
characterized by the presence of two types of magnetically ordered copper
ions whose nuclei experience internal magnetic fields of 64 and 100 kOe. The
changes of the copper NQR spectra described above provide straightforward
evidence that Cu(1) chains that contain no oxygen and belong to the
well-developed Ortho-II phase present an easy diffusion path for the
insertion of water or other related species (e.g. hydroxide ions). The
superconducting volume fraction decreases at this stage approximately at the
same rate as the sharp Cu(1) NQR lines do, but twice as fast as the total
NQR intensity in the 21-33 MHz range (Fig.3). The superconducting volume
fraction for the sample with $z=0.55$ is only 14\% of that for $z=0$. The
simultaneous disappearance of the broad fast-relaxing NQR line at 28 MHz
allows us to assign it to Cu(2) nuclei belonging to the well-developed
Ortho-II phase. It is quite reasonable to suppose \cite{ref3} that water
diffusion is hampered in areas of oxygen-disorder within the crystallites,
i.e. in areas where the empty-chain diffusion channels are blocked by oxygen
in O5 positions, twin boundaries, etc. From this, we assign the ``residual''
NQR spectrum of the $z=0.55$ sample primarily to such areas. A small
increase of the intensity of the terminating Cu(1) NQR line (at 24 MHz) at
the initial stage of the reaction is thus most likely due to the destruction
of long Cu(1) chains.

Our studies of water insertion allow us to tentatively clarify the often
debated question: Do all the nuclei from the superconducting regions of 123
contribute to the NQR/NMR signal, i.e. does the RF fully penetrate
superconducting 123? The observed correlation of the copper NQR intensity,
superconducting volume fraction and water uptake allows us to conclude that 
{\it all} superconducting regions contribute to the NQR signal in the case
of underdoped YBa$_2$Cu$_3$O$_{6.5}$ of 30 $\mu $m-sized powder packed in
Stycast. For comparison, our attempt to insert water (two hours 200$^o$%
C-annealing) into YBa$_2$Cu$_3$O$_7$ lead to a 30\% decrease of the
superconducting volume fraction, but due to better RF field penetration into
a ``spoiled'' superconductor we observed a 20\% {\it increase} of the
observed Cu(2) NQR intensity.

The second stage of the reaction ($z>0.5$) leads to total disappearance of
superconductivity and to dramatic changes in the NQR spectrum (Fig.1). At
maximum water uptake $z_{max}=1.2$ (different from $z_{max}=1$ in Ref.\cite
{ref3}) two new copper centers arise: one characterized by a hyperfine field 
$\sim $80 kOe at the Cu nuclei, and another with a nuclear quadrupole
frequency of 18.4 MHz and a lack of magnetic ordering. It should be
mentioned that magnetically ordered areas which have appeared at $z<0.5$,
are still present and become even larger at $z=1.2$. As estimated by
relative intensities of ZFNMR spectra, they represent about 70-80\% of the
total number of Cu ions participating in well-ordered magnetic phase(s),
which corresponds to about one water molecule ($z\approx 1$) involved in
producing this phase.

X-Ray diffraction spectra taken for the sample with a maximum water uptake
show that the final product of the reaction in our case is identical to the
one considered in Ref.\cite{ref3} as the tetragonal pseudo-248 structure
with 50\% defects at chain copper positions (phase $B$ in Ref.\cite{ref3}).
In this sturcture there are 3 inequivalent Cu sites. We could identify at
least 3 in the magnetically ordered phase, and we find an additional
relatively weak copper NQR line at 18.4 MHz (not far from the 19.8 MHz Cu(1)
NQR line in YBa$_2$Cu$_4$O$_8$). Further studies are necessary to relate the
NMR spectra in detail to the structure.

\section{Conclusions}

Our copper NQR/ZFNMR studies of the reaction of YBa$_2$Cu$_3$O$_{6.5}$
compound with water vapour give straightforward evidence that ``empty'' CuO
chains play the role of easy water insertion channels. The most ordered
regions of the crystallites react most easily. The water insertion reaction
proceeds very slowly at room temperature, but in 6 years in air, water
reaches even samples packed in paraffin \cite{ref1}. At 100-200$^o$C the
reaction proceeds quickly (in few days). The final product of the reaction
is a non-superconducting antiferromagnet characterized by at least two types
of magnetically ordered copper ions with ZFNMR spectra at the frequency
ranges of 46-96 and 96-135 MHz. This antiferromagnetic signal, indicating
decomposition of the superconductor, was even detected in samples packed in
Stycast and left at room temperature (normally deemed a safe storage
procedure) for few years.

\section{Acknowledgments}

This study was supported in part by the Russia Foundation for Basic
Research, under Project 98-02-17687, by INTAS, under Grant 96-0393, and by
the Russian Scientific Council on Superconductivity, under Project 98014.

\section{Figure captions}

Fig.1. Copper NQR spectra for YBa$_2$Cu$_3$O$_{6.5}$ with different water
uptake $z$ (open circles, solid line, solid circles, dotted line and open
triangles correspond to $z=$ 0; 0.14; 0.24; 0.55 and 1.2, respectively): a)
as-taken spectra; b) slow-relaxing part of the spectra (Cu(1) spectra); c)
fast-relaxing part. For details see text.

Fig.2. Copper ZFNMR spectra for YBa$_2$Cu$_3$O$_{6.5}$ with different water
uptake $z$ (open and solid triangles and open and solid circles correspond
to $z=$0.14; 0.24; 0.55 and 1.2, respectively): a) as-taken spectra; b)
spectra normalized by water uptake; the spectrum for $z=1.2$ is normalized
by 0.6.

Fig.3. The dependence of the superconducting volume fraction (solid
circles), copper NQR spectra intensity (solid squares), NQR intensity of
slow-relaxing Cu(1) nuclei (open squares) and ZFNMR intensity (triangles) on
water uptake $z$. Straight lines represent the functions $y=z$ (solid line), 
$y=1-z$ (dashed line) and $y=1-2\cdot z$ (dotted line).

\end{document}